\documentclass{aip-cp}

\usepackage{etoolbox}
\makeatletter
\patchcmd{\@@tablenote}{\xdef}{\protected@xdef}{}{}
\makeatother

\usepackage[numbers]{natbib}
\usepackage{rotating}
\usepackage{graphicx}

\graphicspath{{art/}}

\begin{document}

\title{Searches for Exotic Hadrons at GlueX}

\author[aff1]{Sean Dobbs\corref{cor1}}
\author{the GlueX Collaboration}

\affil[aff1]{Florida State University, 600 W College Ave, Tallahassee, FL 32306}
\corresp[cor1]{Corresponding author: sdobbs@fsu.edu}

\maketitle

\begin{abstract}
The search for hybrid mesons and the detailed study of their spectrum is the primary goal of the GlueX Experiment in Hall D at Jefferson Lab.  The identification and study of hybrid mesons promises to provide unique insight into gluonic degrees of freedom in QCD and the nature of confinement. The experiment combines an intense photon beam with linear polarization peaking around 9 GeV incident on a liquid hydrogen target with a nearly hermetic spectrometer, allowing for the comprehensive study of charged and neutral particle final states. The first phase of the experiment has recently concluded, yielding a photoproduction data set of unprecedented size and quality. We report on the status of the analysis of this data, including measurements of polarization observables, progress in spectroscopic measurements of light mesons, and the measurement of the photoproduction of $J/\psi$ near threshold, which is providing critical insight into the nature of the LHCb closed-charm pentaquark candidates.
\end{abstract}

\section{INTRODUCTION}

The search for exotic hadrons beyond the standard $q\bar{q}$ mesons and $qqq$ baryons has been given new life in recent years with the advent of modern, high-luminosity experiments which have led to a series of exciting new discoveries and possibilities.  For example, in the charm quark sector, dozens of new, often unexpected states have been identified, starting with the so-called X(3872) (recently renamed the $\chi_{c1}(3872)$ by the PDG~\cite{pdg}).  Many of these states do not fit within the standard quark model and a variety of proposals exist, but the most provocative of these are the charged exotic states, whose decays strong suggest a 4 quark ($Z_c$) or 5 quark ($P_c$) nature of some variety.  There are several different additional measurements that can be made to distinguish between these models, and a powerful method is to search for their production through processes other than that in which they were originally identified.  The creation of new photoproduction facilities of the right energy range has opened new opportunities for such measurements.

An even more exciting possibility is the identification and study of hadrons in which the excitation of their confining gluonic field directly contributes to the properties of these states~\cite{hybrids}.   In the meson sector, the existence of such  ``hybrid'' mesons is indicated through calculations of QCD on the lattice (LQCD)~\cite{lattice}, and their identification and study would yield a new way to study the properties of the gluonic field.  The GlueX Experiment at Jefferson Lab in has recently been built to search for these states, and has been producing physics data since 2017.  The experiment combines an intense photon beam with a large acceptance, general purpose spectrometer, which also provides the ability to address a wide range of questions in hadronic physics.

Hybrid mesons are generally expected to be found in the $1.5-2.5$~GeV mass region, and are expected to have both the same $J^{PC}$ quantum numbers as normal mesons ($J^{PC} = 0^{-+}, ~1^{--},~2^{-+}$) and exotic quantum numbers which are forbidden to normal mesons ($J^{PC} = 0^{+-}, ~1^{-+},~2^{+-}$).  Although looking for exotic quantum number mesons is a natural place to start the search, it is important to identify normal and exotic quantum number hybrid mesons in multiple decays modes, in order to determine the nature of these states and thereby the properties of the gluonic excitations. Because of the many wide, overlapping mesons which decay into the same final states in this mass region, amplitude analysis of the data is needed to extract resonance properties from experiment.

There is a long history of searches for hybrid mesons~\cite{hybrids}, mostly focused on the search for the light exotic $1^{-+}$ state, the $\pi_1$.  Most searches have been in pion production, with the strongest evidence for the $\pi_1$ being in its decays to $\pi\eta'$ and $\rho\pi$.  The most precise data in these modes is from the COMPASS experiment, and the strongest evidence for the $\pi_1$ is from a coupled channel fit to the published COMPASS data~\cite{compassetapi} by the JPAC Collaboration~\cite{jpacetapi}.  With their unitary reaction model, they are able to consistently describe both $\pi\eta$ and $\pi\eta'$ spectra with one $\pi_1$ resonance with parameters $M=1564\pm24(\mathrm{stat})\pm86(\mathrm{syst})$~MeV and $\Gamma=492\pm54(\mathrm{stat})\pm102$~MeV.
While this is a major step forward in resolving the ambiguity in the number of $\pi_1$ states claimed in these reactions, to more firmly determine the identification of this state it is desirable to identify it at a similar level in a different reaction.

Photoproduction is a particularly exciting process to search for the production of such hybrids.  The photon-proton interaction can produce any expected hybrid state through Vector Meson Dominance, and there is little existing data at photon beam energies of $E_\gamma \approx 9$~GeV, which is ideal for searching in the hybrid meson mass range in a fixed target experiment, particularly in final states containing neutral particles.  Additionally, if the spin-1 photon is polarized, this can be used to further constrain production processes in amplitude analyses.  Performing such photoproduction measurements is therefore an appealing prospect, and the GlueX Experiment was designed and built to perform them.

\section{THE GLUEX EXPERIMENT AND THE PATH TOWARDS HYBRID MESONS}

\begin{figure}[!tb]
\includegraphics[width=5.5in]{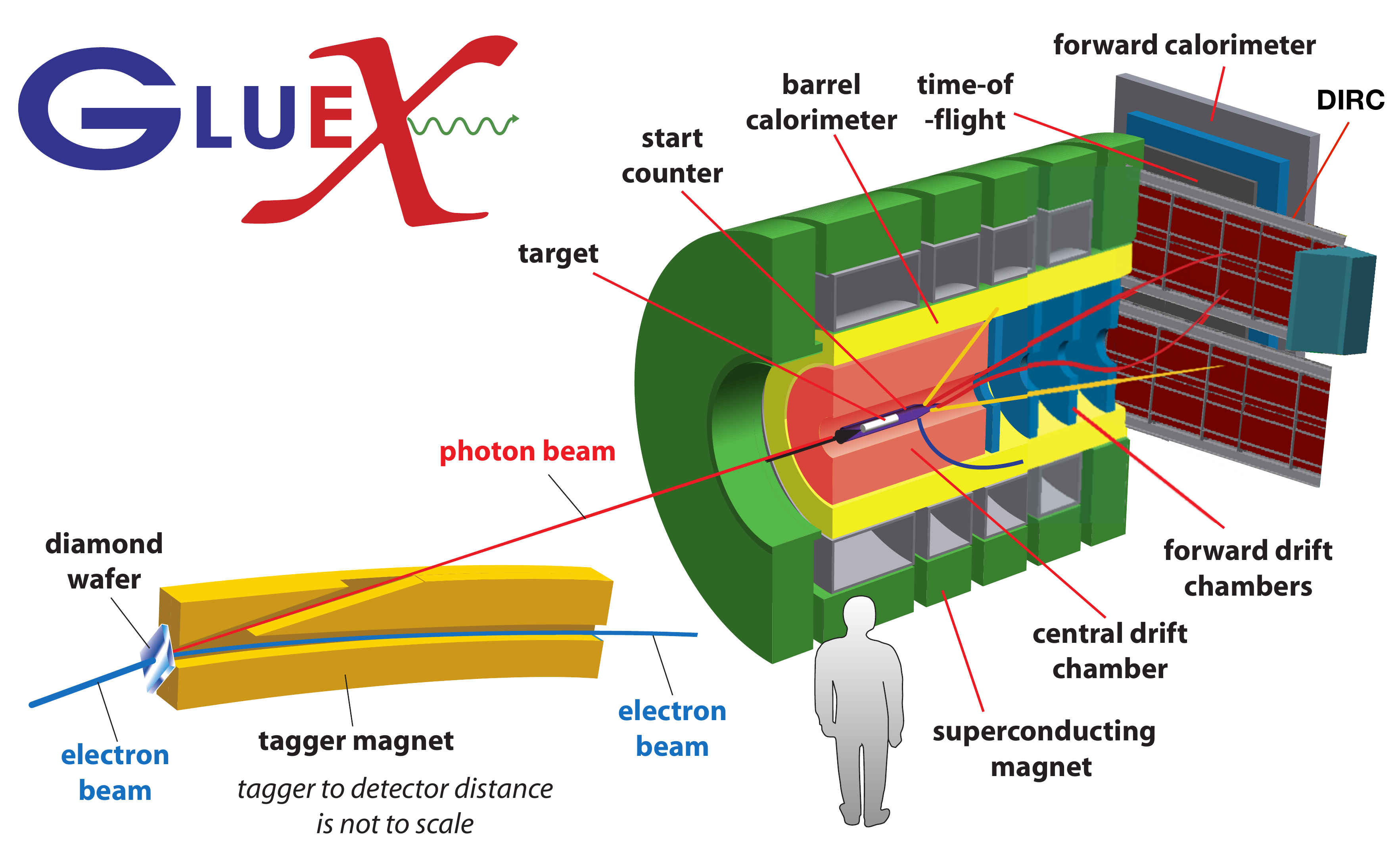}

\caption{Schematic of the GlueX detector highlighting the major components.  The DIRC was installed after the first physics run and is being commissioned.}
\label{fig:detector}

\end{figure}

The GlueX Experiment combines a solenoidal spectrometer with almost full acceptance for charged and neutral particles with an intense photon beam that has a high degree of linear polarization, and is schematically illustrated in Fig.~\ref{fig:detector}.  The experiment has been described elsewhere in detail~\cite{gluex-pi0,gluex-jpsi,gluex-hadron}, which we briefly summarize in the following.  The electron beam from the CEBAF accelerator strikes a thin diamond and produces a beam of photons through bremsstrahlung, including a component with a high degree of linear polarization due to coherent bremsstrahlung.  The enhancement in photon flux due to this coherent process peaks at a beam energy of 9~GeV with an average polarization of around 40\%.  
The photon energies are determined by tagging the scattered electrons with a dipole spectrometer that has coarse and finely segmented scintillators yielding tagged photon energy resolutions of $E_\gamma \approx 10-25$~MeV.
The photon beam propagates 75~m, after which it is collimated, and its flux is measured through the rate of $\gamma\to e^+e^-$ conversions in a two-armed spectrometer~\cite{gluexps}.  The flux is $1-5\times10^7~\gamma/s$ in the peak. The photon beam polarization is measured through the scattering of photons off of atomic electrons, $\gamma e^- \to e^-e^+e^-$, where the scattered electron is measured in a small silicon detector~\cite{gluextpol}.

The photon beam is incident on a 30~cm~LH$_2$ target, which sits inside a solenoid magnet that can reach a magnetic field of 2T.  The target is surrounded by a Start Counter~\cite{gluexsc} of bent scintillator paddles, used to select photon bunches for particle identification, the straw-tube-based Central Drift Chamber~\cite{gluexcdc}, and the lead-scintillator-matrix Barrel Calorimeter~\cite{gluexbcal}.  In the forward direction, the Forward Drift Chamber~\cite{gluexfdc} lies inside of the bore of the magnet, and consists of 4 packages of drift chambers. Outside of the magnet are placed a pair of Time-of-Flight scintillator paddle walls and a Forward Calorimeter~\cite{gluexfcal} made of lead-glass crystals.  This combination of detectors allows for reconstruction of charged and neutral particles over almost the full solid angle with nearly uniform azimuthal efficiency.  The momentum resolution for charged particles is $\sigma(p)/p \approx 1-5\%$, where the resolution drops as the particles increase in momentum and decrease in polar angle, and the energy resolution for photons is  $\sigma(E)/E \approx 6\%/\sqrt{E} + 2\%$.
After a brief commissioning run in 2016, the first full physics run commenced in 2017 and was finished by the end of 2018.  The 2017 data correspond to 20\% of the total, are fully reconstructed and calibrated, and are used for most of the results shown herein, while the processing of the 2018 data is well underway and is expected to be finished by this Fall.  A DIRC detector for enhanced particle identification in the forward direction has been installed, and will be fully commissioned by the end of the year.

\subsection{Searching for Exotics in Photoproduction}

A detailed understanding of the light meson spectrum in photoproduction requires amplitude analysis in order to disentangle the many hadronic resonances, which are often broad and overlapping, that contribute to the final states under investigation.  Models for the production and decay of these resonances are required for such an analysis, and developing these requires close collaboration between experimentalists and theorists. 

In order to constrain models of meson photoproduction, we can take advantage of the polarized GlueX photon beam. A program of measuring the $\Sigma$ beam asymmetry in single-pseudoscalar-meson photoproduction reactions, such as $\gamma p \to (\pi^0,\eta,\eta') p$ and $\gamma p\to (\Lambda^0,\Sigma^0) K^+$, and the spin-density matrix elements in single-vector-meson photoproduction reactions, such as $\gamma p \to (\rho^0,\omega,\phi) p$, is underway.  Many of these analyses are discussed in detail elsewhere in these Proceedings~\cite{aaustreg,sfegan,wmcginley,nilanga}, and have already lead to one published paper~\cite{gluex-pi0} and three others which are currently undergoing internal review.

The next step is to measure production cross sections, whose dependence on photon beam energy and Mandalstam-t yields more information to inform production models.  To measure these cross sections, a firm understanding of such aspects as experimental acceptance, reconstruction efficiency, and photon flux is required.  A systematic program of measuring $d\sigma/dt$ of single pseudoscalar and vector mesons for $E_\gamma > 3$~GeV is underway.  This will allow comparison with previously published results and extend our understanding of the evolution of production processes from the nucleon-resonance-dominated regime to the higher energies where the meson production dominates.  Other cross section measurements are in progress as well. For example, a program of studying the photoproduction of Cascade hyperons is underway, with the first preliminary results of the ground state $\Xi^-$ photoproduction cross section reported elsewhere in these Proceedings~\cite{acernst}.

Once the requisite understanding of the detector and the underlying production processes has been established, amplitude analyses will be performed to study the ``normal'' mesons that can be found in photoproduction and to search for evidence of the ``hybrid'' mesons.  Initial efforts have begun in this direction, and below I illustrate the prospect for some early measurements.

The large size and high quality of the data collected by the multipurpose GlueX detector also allows for new ideas and other opportunistic measurements.  Two exciting new developments include the study of $\Lambda\overline{\Lambda}$ photoproduction, described elsewhere in these Proceedings~\cite{hao}, and the measurement of $J/\psi$ photoproduction, which I describe below.

\begin{figure}[!tb]
\includegraphics[width=2.2in]{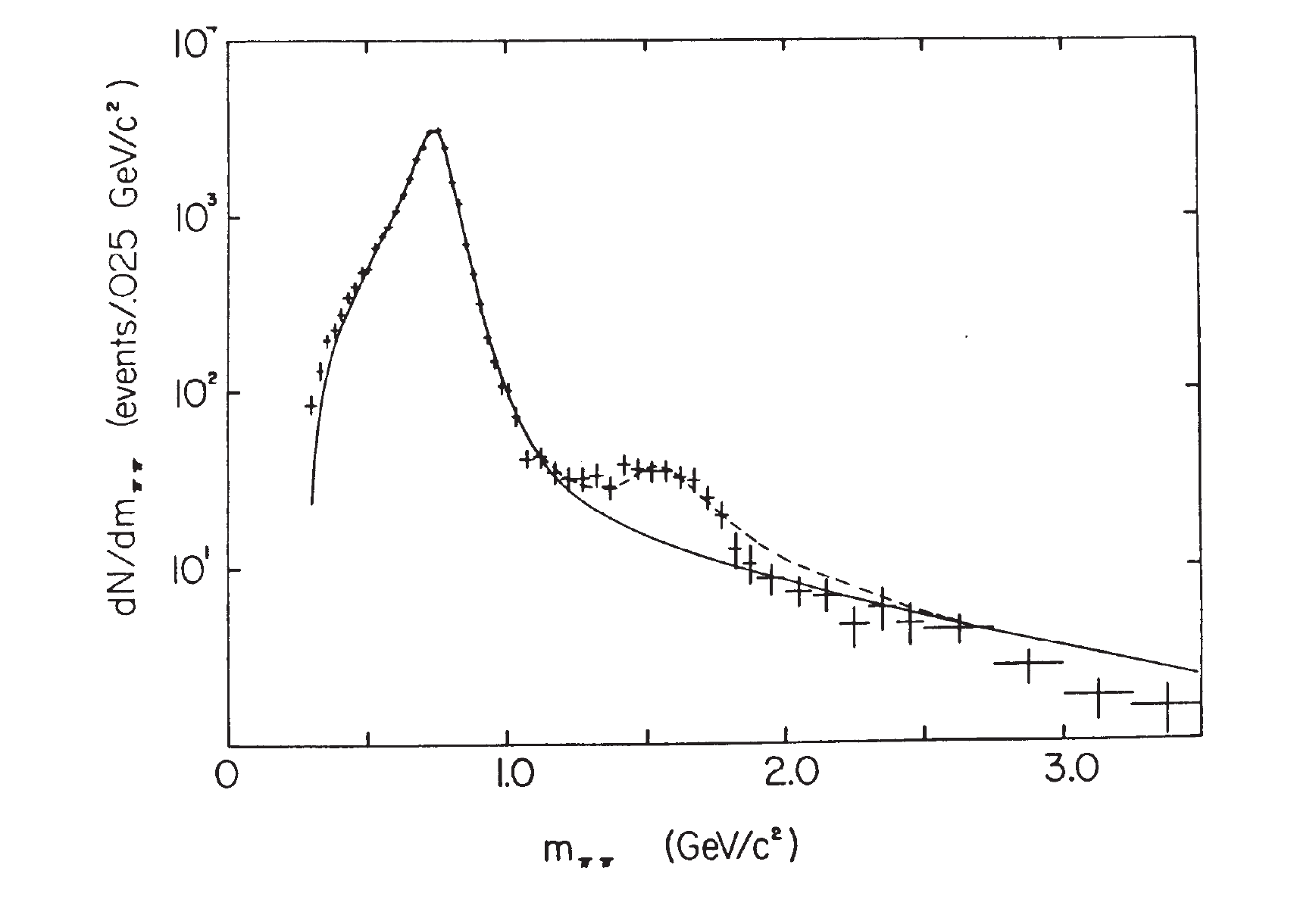}
\raisebox{12pt}{\includegraphics[width=1.9in]{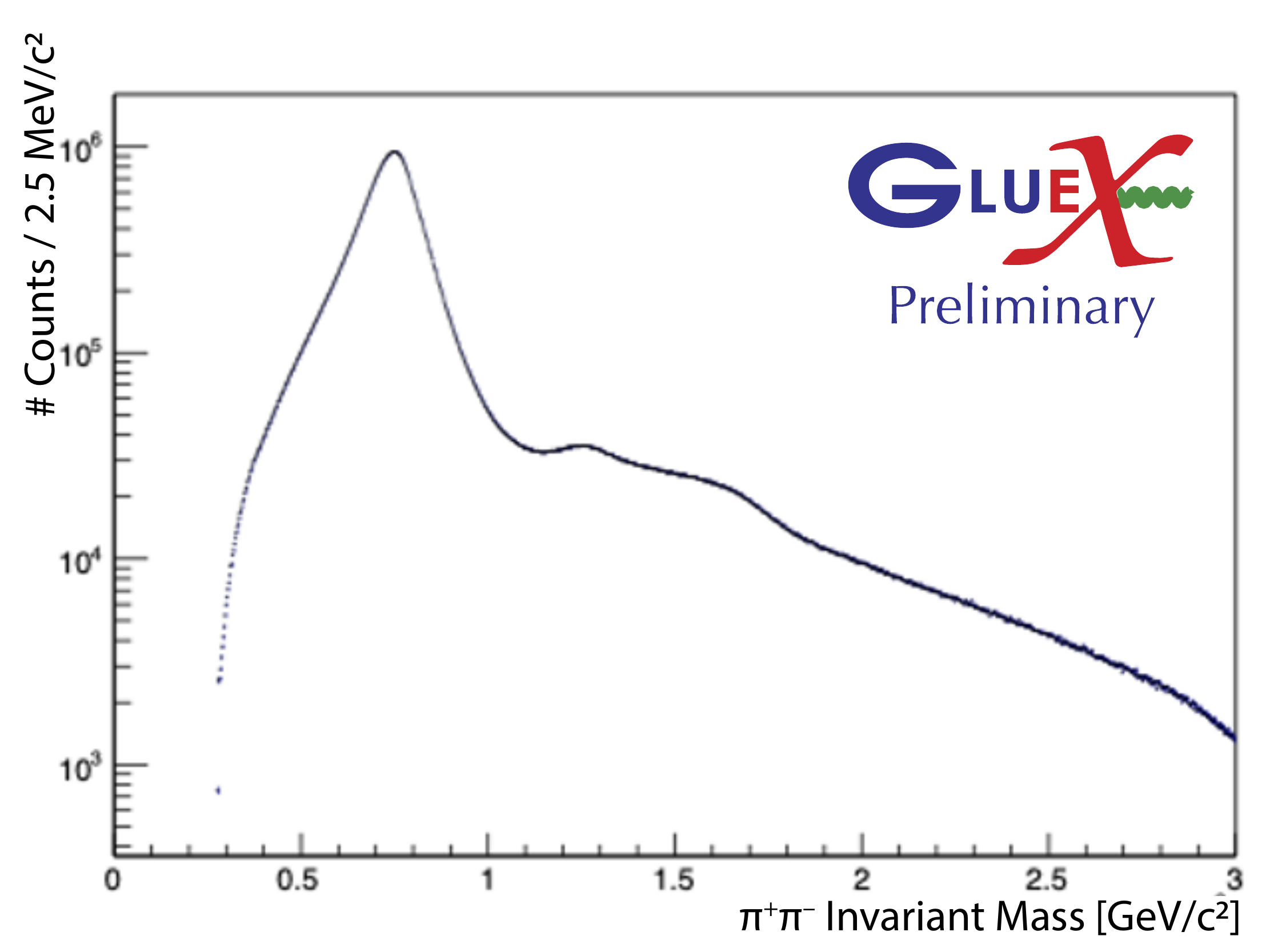}}
\raisebox{8pt}{\includegraphics[width=1.9in]{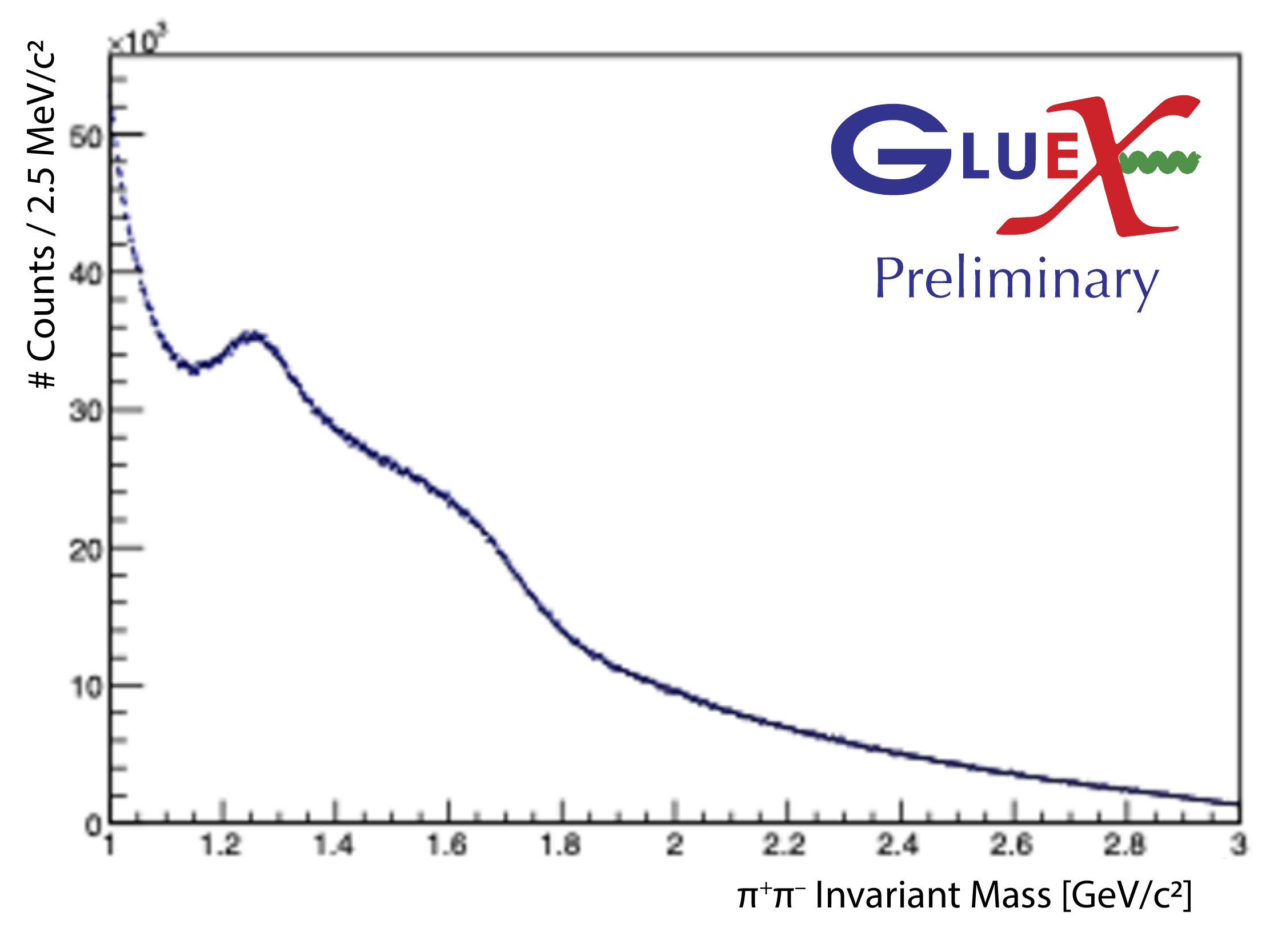}}

\caption{Dipion invariant mass distributions from reconstructed $\gamma p \to \pi^+\pi^- p$ events: (Left) SLAC 20~GeV photon beam~\cite{slac-pipi}; (Middle) Very preliminary results from GlueX 9~GeV photon beam; (Right) The distribution in the middle panel but excluding most of the $\rho$ mass region.}
\label{fig:pipimass}

\end{figure}

\subsection{Early Spectroscopy Prospects}

The path to establishing the full spectrum of hybrid mesons continues to be a long path, and a natural first step for GlueX is to look for the spin-1 hybrids in a simple set of decays, specifically to two pseudoscalar mesons.  For the $1^{-+}$ exotic hybrids, this would allow for the establishment of the $\pi(1600)$ in its decays to $\eta\pi$ and $\eta'\pi$, and the search for its partners the $\eta_1$ and $\eta_1'$ in their decays to $\eta\eta'$~\cite{hybrids}. For the $1^{--}$ vector hybrids, the first goal would be to firmly establish the spectrum of excited $\rho$ and $\phi$ mesons and then look for deviation from quark model or LQCD predictions.

The status of the excited vector $\rho$ meson spectrum remains unsettled, with the PDG listing $\rho(1450)$ and $\rho(1700)$ as well-established states but with their identities not firmly established~\cite{pdg-rho}.  Additional precision data on these resonances is required, particularly in the mass region $>1.6$~GeV, which fortuitously is the mass range that GlueX is designed to study.  Additionally, information of the photoproduction of these states should yield helpful information complementary to that obtained from other production processes.  In a previous measurement of the reaction $\gamma p \to \pi^+\pi^- p$ with a 20~GeV photon beam at SLAC, a clear, wide enhancement in the $\pi^+\pi^-$ invariant mass spectrum peaking around 1.6~GeV was seen~\cite{slac-pipi}.  We have looked at the same reaction at GlueX, and using the data available for analysis find that our statistics are two orders of magnitude larger than those of the SLAC measurements, and are of sufficient precision to show two additional bumps at higher mass, likely due to the $f_2(1270)$ and an excited $\rho$ meson, as illustrated in Fig.~\ref{fig:pipimass}.  Amplitude analysis of the data in order to determine the resonance contribution to the mass spectrum is underway, which will provide a better determination.  Analysis of charged and neutral kaon pair production is also underway to search for excited $\phi$ states.

\begin{figure}[!tb]
\begin{tabular}{cc}
\raisebox{12pt}{\includegraphics[width=2.9in]{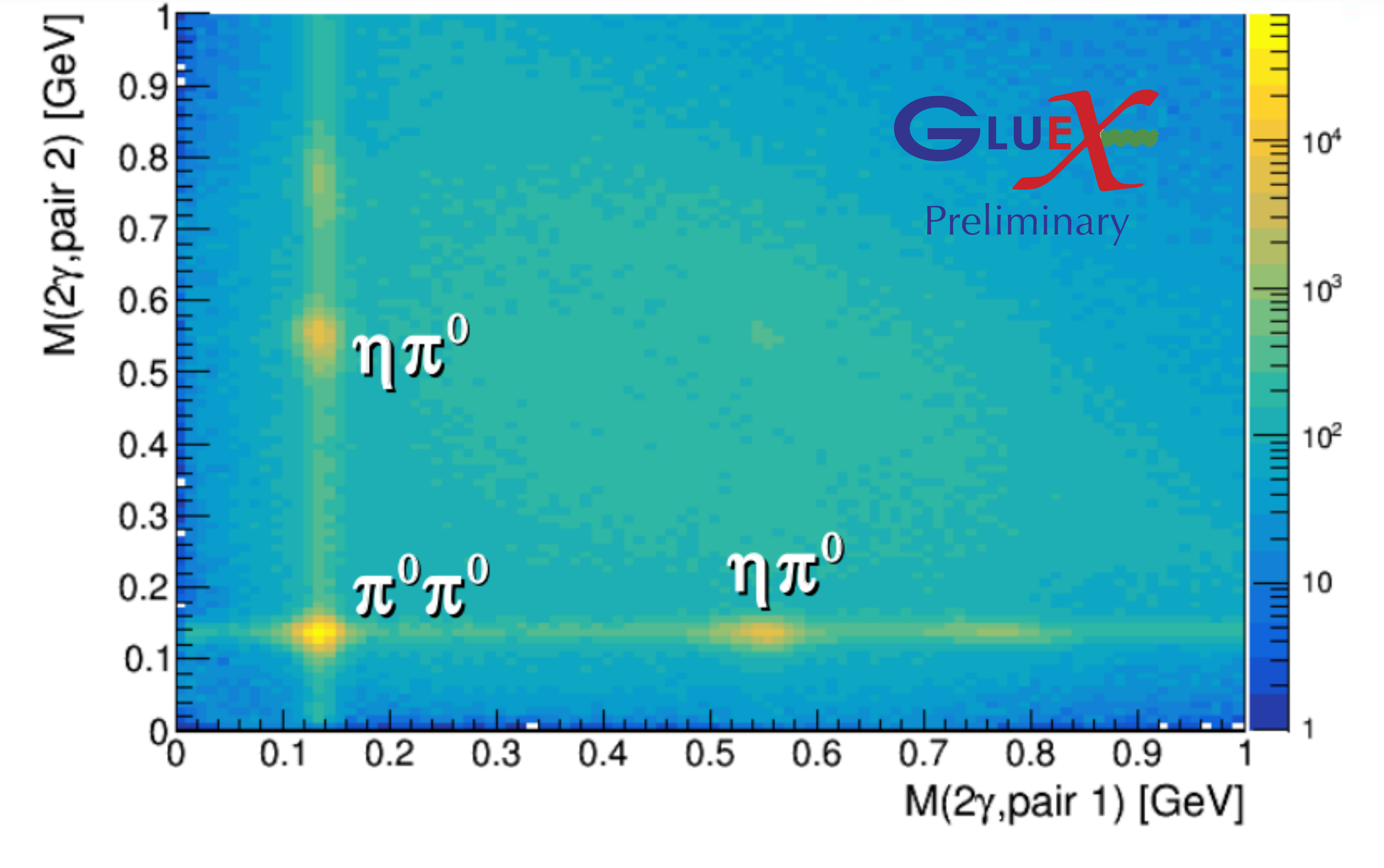}} &
\includegraphics[width=3.2in]{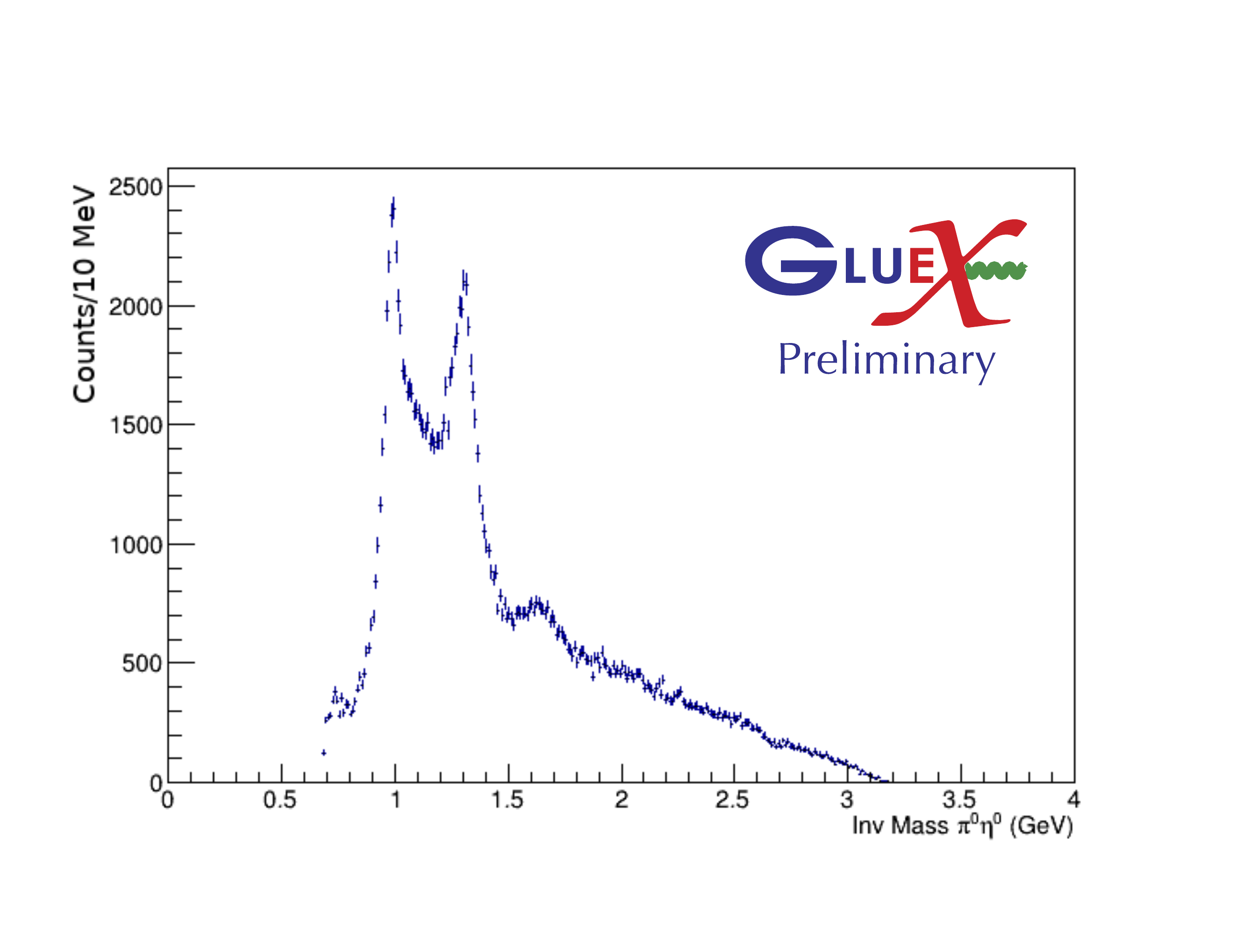} \\

\raisebox{5pt}{\includegraphics[width=2.8in]{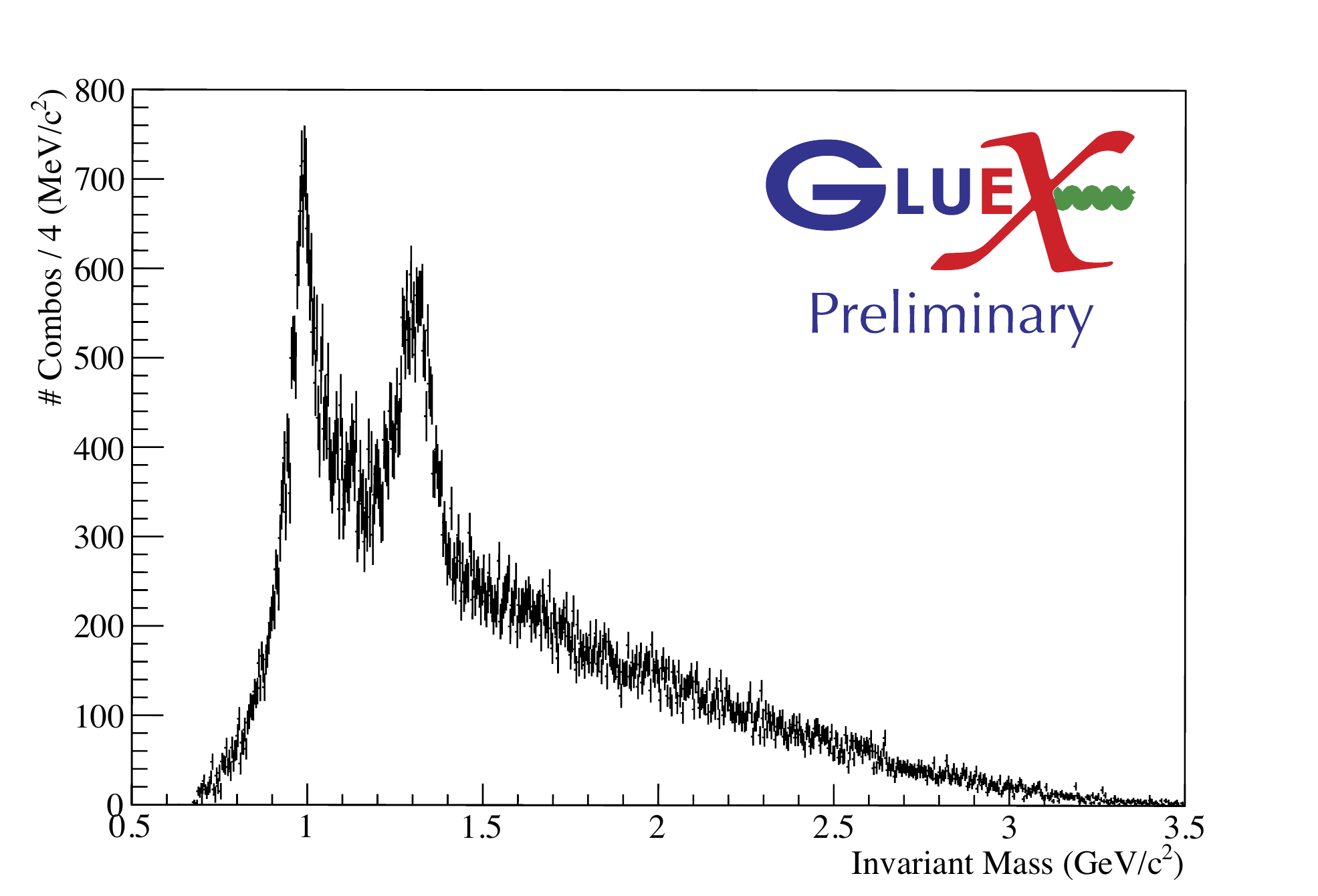}} &
\includegraphics[width=3.in]{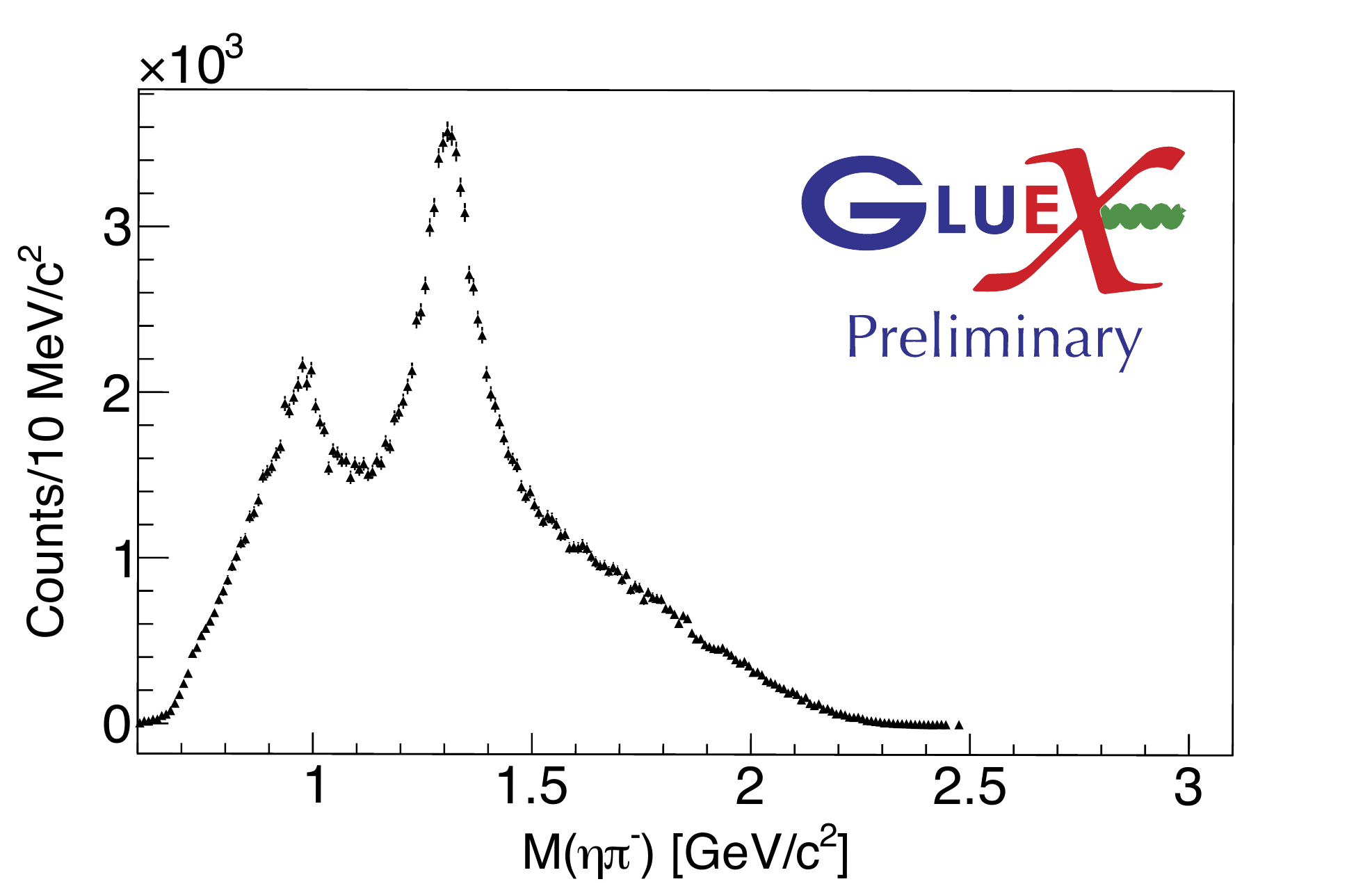}
\end{tabular} 

\caption{Preliminary $\pi\eta$ mass spectra from GlueX data: (Top left) 2D plot of diphoton invariant mass combinations from a subset of existing $\gamma p \to 4\gamma p$ data, showing clear signals from $\pi^0\eta \to 4\gamma$ events.  $\eta\pi$ invariant mass distributions from: (Top right) $\gamma p \to \pi^0 \eta$,  $\eta\to\gamma\gamma$; (Bottom left) $\gamma p \to \pi^0 \eta$,  $\eta\to\pi^+\pi^-\pi^0$; (Bottom right)  $\gamma p \to \Delta^{++} \pi^- \eta$.}
\label{fig:pietamass}

\end{figure}

As previously discussed, the $\eta\pi$ and $\eta'\pi$ channels are important for the study of the $\pi_1(1600)$ meson, and they have been an early focus of GlueX analysis. As illustrated in Fig.~\ref{fig:pietamass}, we have begun the study of the reaction $\gamma p \to \pi^0 \eta$, where the $\eta$ is reconstructed in the $\eta\to\gamma\gamma$ and $\eta\to\pi^+\pi^-\pi^0$ channels.  In both cases, several structures are seen, notably strong excitations identified with the well-known $a_0(980)$ and $a_2(1320)$ mesons.  We have also begun the study of the isospin-partner $\pi^-\eta$ channel with the reaction $\gamma p \to \Delta^{++} \pi^- \eta$, $\eta\to\gamma\gamma$, as shown in Fig.~\ref{fig:pietamass}.
From these studies, we expect that the full GlueX data currently on tape should correspond to roughly twice the statistics of the COMPASS measurements in these channels, and the prospects for identifying the $\pi_1(1600)$ production are strong.

\begin{figure}[!tb]

\includegraphics[width=5.5in]{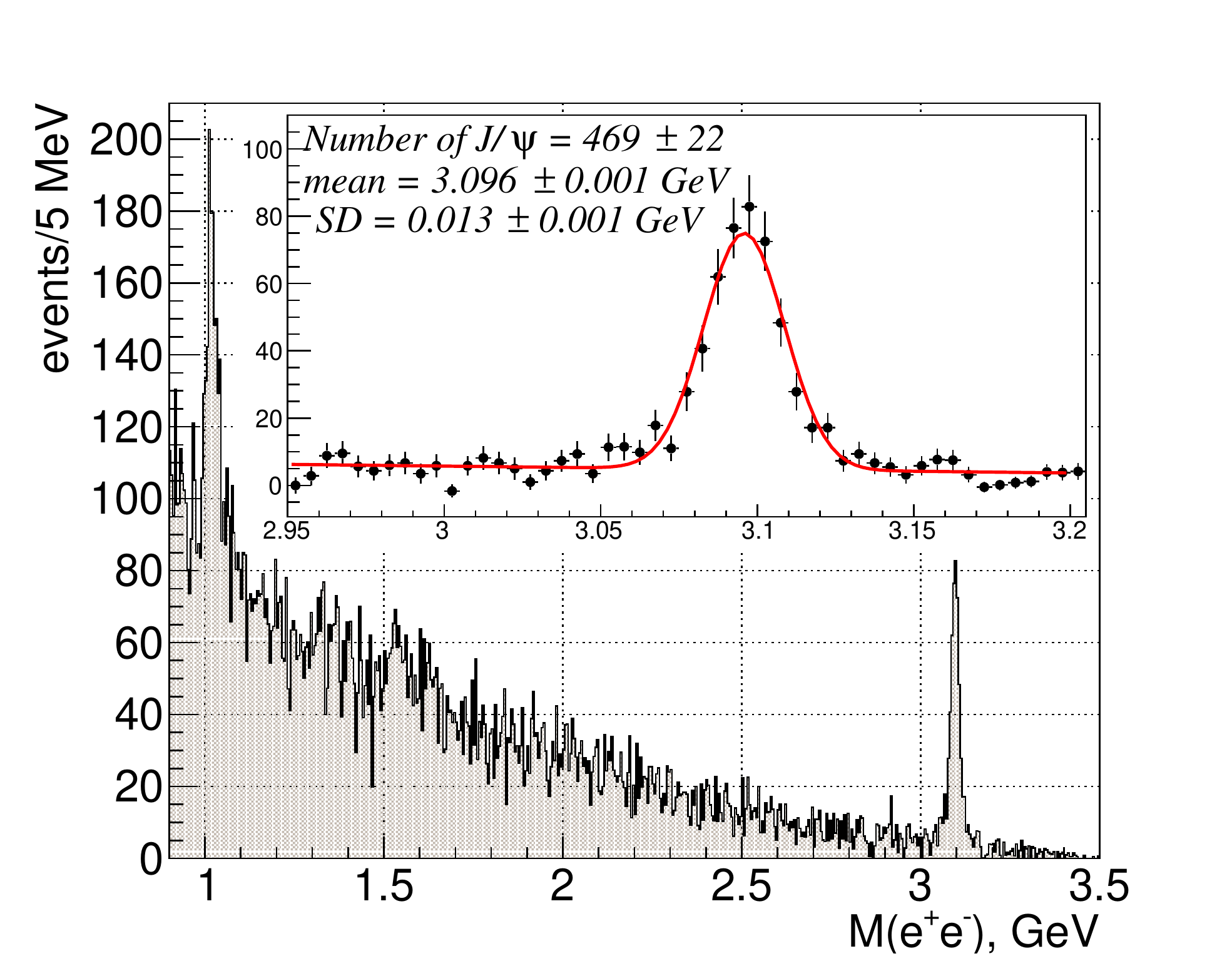}

\caption{Electron-positron invariant mass spectrum from GlueX data~\cite{gluex-jpsi}. The insert shows the $J/\psi$ region fitted with the sum of a Gaussian and a linear polynomial.}
\label{fig:jpsimass}
\end{figure}

\begin{figure}[!tb]

\includegraphics[width=5.5in]{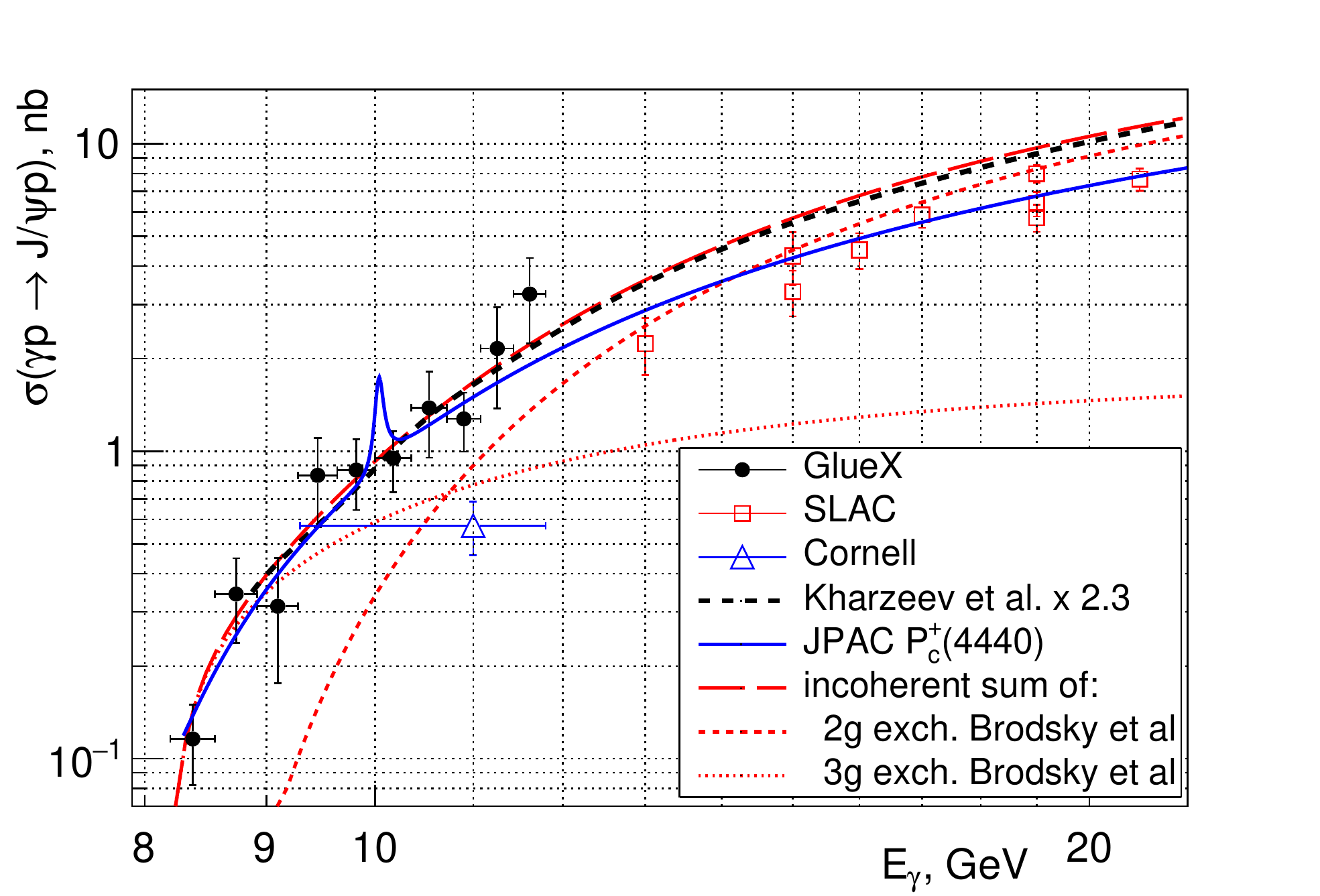}

\caption{Total cross section measurements for $J/\psi$ photoproduction, compared with previous measurements and theoretical models (from Ref.~\cite{gluex-jpsi}).}
\label{fig:jpsixsec}
\end{figure}

\section{CHARMONIUM PRODUCTION NEAR THRESHOLD}

The photon beam energies available to GlueX allow the study of the production of the bound charmonium ($c\bar{c}$)  states, from the ground state $\eta_c(1S)$ at a threshold of $E_\gamma=7.7$~GeV, to the radial excitation of the vector state $\psi(2S)$ at a threshold of $E_\gamma=10.9$~GeV. Furthermore, the large acceptance of the spectrometer make it an ideal experiment to measure the wide range of exclusive reactions required to identify these states.  The first step in these studies is to establish the production of $J/\psi(1S)$, which has a threshold of $E_\gamma=8.2$~GeV.  There is little existing data on this reaction near threshold, with just two experiments from the 1970s that were not fully exclusive and performed on nuclear targets.

Charmonia primarily interact with the light quarks of the nucleon by exchanging gluons, so the exclusive production of these states near threshold provides an excellent tool for studying the gluonic field in the nucleon.  The energy dependence of the total cross section for elastic $J/\psi$ photoproduction has been studied in several theoretical frameworks.  This energy dependence has been evaluated using dimensional scaling rules where different shapes are found depending on the number of gluons exchanged in the interaction~\cite{brodsky}.  In particular, very close to threshold it is expected that a contribution may be found from the case where all three valence quarks in the nucleon exchange high-x gluons with the $c\bar{c}$ system.  Several other calculations have found that the shape of the energy dependence of this cross section depends on the contribution of gluons to the nucleon mass, particularly through the trace anomaly term of the QCD energy-momentum tensor~\cite{kharseev,hatta}.  

The reactions discussed so far have been in the $t-$channel, but if we look at the $s-$channel, there is the exciting prospect of directly producing 5-quark states, with a quark content of $c\bar{c}uud$.  Observations of such states have been reported by the LHCb experiment in the study of the decay $\Lambda_b^0 \to J/\psi p K^-$. The LHCb Collaboration initially reported the identification of two states~\cite{lhcb1,lhcb2}, the broad $P_c^+(4380)$ and the narrow $P_c^+(4450)$, with preferred spin assignments of one with spin--3/2 and the other with spin--5/2 and opposite parity.  More recently~\cite{lhcb3}, they reported the identification of three narrow states, with the $P_c^+(4450)$, splitting into the  $P_c^+(4440)$ and $P_c^+(4457)$ and a new $P_c^+(4312)$ appearing, although without spin-parity assignments as the full partial wave analysis has not been performed yet.  It was soon realized that these states could also be produced in the $s-$channel photoproduction reaction, $\gamma p \to P_c^+ \to J/\psi p$~\cite{wang,voloshin,karliner,jpac}, which would provide a complementary measurement which is free from three-body rescattering effects that could exist in the $\Lambda_b^0$ decays~\cite{triangle}.  Assuming this reaction can be described using Vector Meson Dominance, only one free parameter describes the production of the $P_c^+$ states, the branching fraction $\mathcal{B}(P_c^+ \to J/\psi p)$.

We have studied the process $\gamma p \to J/\psi p$, $J/\psi \to e^+e^-$ using data from the commissioning run, and about 20\% of the current physics run.  We reconstruct the proton and electron-positron pair, identifying electrons using information from the calorimeter, which results in a background suppression factor of $\approx5000$, and applying a kinematic fit to impose energy-momentum conservation and improve the reconstructed mass resolution.  The resulting spectrum is shown in Fig.~\ref{fig:jpsimass}, where clear peaks corresponding to $\phi\to e^+e^-$ and $J/\psi \to e^+e^-$ are seen.  The region in between the two peaks has contributions from non-resonant $\gamma p \to e^+e^- p$ production, and reactions with misidentified pions.  The fit shown in the insert of Fig.~\ref{fig:jpsimass} yields a total of $N(J/\psi) = 469\pm22$ for the data under analysis, with a mass resolution of $\sigma(J/\psi)=13$~MeV.  

The absolute efficiencies of the GlueX spectrometer are still under study, but we can also measure the $J/\psi$ cross section relative to that of the non-resonant $\gamma p \to e^+e^- p$, which can be calculated to a good accuracy using standard QED calculations.  The non-resonant yields are extracted from the residual pion background by fitting the shape of the $E(\mathrm{calorimeter})/p(\mathrm{track})$ distribution used for electron ID.  The analysis is performed in bins of the beam energy, and the total cross section is determined using the measured yields, the calculated non-resonant cross-section, and the Monte Carlo-determined relative efficiency between the two processes.  

The cross-sections from our data are shown in Fig.~\ref{fig:jpsixsec}, along with the previous measurements in this energy range and various model calculations.  At their present level of precision, we find the shape of our data is generally consistent with these models, when a substantial contribution from 3-gluon exchange is assumed in the model of Ref.~\cite{brodsky}.  Additionally, we see no clear contribution from the $P_c^+$ states.  We determine upper limits including systematic uncertainties for $\mathcal{B}(P_c^+ \to J/\psi p)$ using a variant of the model in Ref.~\cite{jpac} assuming $J^P(P_c^+)$ at 90\% confidence level of 4.6\%, 2.3\%, and 3.8\% for $P_c^+(4312)$, $P_c^+(4440)$, and $P_c^+(4457)$, respectively.  Due to the assumption of vector meson dominance, these upper limits will scale roughly as a factor of $(2J+1)$.  
We also set less-model-dependent limits on $\sigma_\mathrm{peak}(\gamma p \to P_c^+)\times\mathcal{B}(P_c^+ \to J/\psi p)$, using an incoherent sum of a Breit-Wigner and the non-resonant component of the model of Ref.~\cite{jpac}. We find upper limits on this quantity at 90\% confidence level of 4.6, 1.8, and 3.9~nb for $P_c^+(4312)$, $P_c^+(4440)$, and $P_c^+(4457)$, respectively.

Since the conference, the analysis has been published in Physical Review Letters~\cite{gluex-jpsi}.  The full data on tape is currently under analysis, which is expected to correspond to over 1500 $J/\psi$'s.  This data will allow for a detailed measurement of the cross-section as a function of $t$ and beam energy simultaneously, as well as searches for other charmonium states.  Unbinned fits to this data are expected to provide better sensitivity for the extraction of t-channel exchange parameters and the search for $P_c$ states.

\section{SUMMARY}

The GlueX experiment finished its first phase of data taking by the end of 2018, and the data set of unprecedented size which has been collected is enabling a wide-ranging physics program.  As the first steps towards the amplitude analyses required for the identification of the spectrum of hybrid mesons, a broad program of polarization observable, spin-density matrix element, and cross section measurements is under way.   Initial invariant mass spectra show that the GlueX data are several orders of magnitude larger than previous photoproduction experiment in the same beam energy region, and the number of events in final states relevant for hybrid meson searches are of the same order of magnitude or larger than existing pion beam experiments.  This data also enables measurements of other rare reactions, such as near-threshold charmonium production.  Using a portion of the total data collected, we have measured the total $J/\psi$ photoproduction cross section dependence on the incident photon energy, which is an important input into models of the gluonic structure of the nucleon and allow us to probe the photoproduction of the LHCb $P_c$ states.  We place a model-dependent upper limit on $\mathcal{B}(P_c^+ \to J/\psi p)$ of less than a few percent, which already excludes some molecular models of the internal structure of the $P_c$'s.
The next phase of the GlueX experiment will begin in Fall 2019~\cite{jrstevens}, featuring a higher photon beam intensity and enhanced pion/kaon separatation.  This phase is expected to collect roughly a factor 5 larger data set than currently exists, and will also allow for a stronger focus on hadrons containing strange quarks.




\section{ACKNOWLEDGMENTS}
This material is based upon work supported by the U.S. Department of Energy, Office of Science, Office of Nuclear Physics under contract DE-AC05-06OR23177. The author acknowledges the support of Jefferson Science Associates, LLC.


\nocite{*}
\bibliographystyle{aipnum-cp}%

\end{document}